\documentclass[aps,prd,twocolumn,groupedaddress,showpacs]{revtex4}
\usepackage{epsf,epsfig,graphicx}
\begin{document}

\title{Naturally large tensor-to-scalar ratio in inflation}

\author{Guo-Chin Liu$^{1}$, Kin-Wang Ng$^{2,3}$, and I-Chin Wang$^{1}$}

\affiliation{
$^1$Department of Physics, Tamkang University, Tamsui, New Taipei City 25137, Taiwan\\
$^2$Institute of Physics, Academia Sinica, Taipei 11529, Taiwan\\
$^3$Institute of Astronomy and Astrophysics, Academia Sinica, Taipei 11529, Taiwan
}

\vspace*{0.6 cm}
\date{\today}
\vspace*{1.2 cm}

\begin{abstract}
Recently, BICEP2 measurements of the cosmic microwave background (CMB) $B$-mode polarization at degree angular scales has indicated the presence of tensor modes with a high tensor-to-scalar ratio of $r=0.2$ when assuming nearly scale-invariant tensor and scalar spectra, although the signal may be contaminated by dust emission as implied by the recent {\em Planck} polarization data. This result is in conflict with the {\em Planck} best-fit Lambda Cold Dark Model with $r<0.11$. Due to the fact that inflaton has to be interacting with other fields so as to convert its potential energy into radiation to reheat the Universe, the interacting inflaton may result in a suppression of the scalar spectrum at large scales. This suppression has been used to explain the observed low quadrupole in the CMB anisotropy. In this paper, we show that a combination of the tensor modes measured by BICEP2 and the large-scale suppressed scalar modes contributes to the CMB anisotropy in such a way that the resultant CMB anisotropy and polarization power spectra are consistent with both {\em Planck} and BICEP2 data. We also project our findings to cases in which $r$ may become reduced in future CMB polarization measurements.
\end{abstract}

\maketitle

\section{Introduction}

The spatial flatness and homogeneity of the present
Universe strongly suggest that a period of de Sitter expansion
or inflation had occurred in the early Universe~\cite{olive}.
During inflation, quantum fluctuations of the inflaton
field may give rise to energy density perturbations (scalar
modes)~\cite{pi}, which can serve as the seeds for the formation
of large-scale structures of the Universe. In addition,
a spectrum of gravitational waves (tensor modes) is produced
from the de Sitter vacuum~\cite{star}.

In the standard big bang cosmology, scalar and tensor modes leave signatures
on the cosmic microwave background (CMB) thoroughly determined by the power spectra, $C_l^{TT}$,
$C_l^{TE}$, $C_l^{EE}$, and $C_l^{BB}$, where $T$, $E$, and $B$
denote the temperature anisotropy, $E$-mode polarization, and $B$-mode polarization, respectively.
CMB anisotropy and $E$-mode polarization have been well measured by WMAP, {\em Planck}, and many other
experiments~\cite{planck} (and references therein). Unlike scalar modes,
tensor modes are very weakly coupled to matter, so once produced
they remain as a stochastic background until today. However, they induce large-scale CMB
anisotropy via the Sachs-Wolfe effect and uniquely $B$-mode polarization.
Detecting these signals would provide a potentially important probe of the inflationary epoch
and the latter is the primary goal of ongoing and future CMB experiments~\cite{weiss}.
Recently, WMAP+SPT CMB data has placed an upper limit on
the contribution of tensor modes to the CMB anisotropy, in
terms of the tensor-to-scalar ratio, which is $r<0.18$ at $95\%$ confidence level, tightening to $r<0.11$
when also including measurements of the Hubble constant
and baryon acoustic oscillations (BAO)~\cite{story}. Planck
Collaboration XVI has quoted $r<0.11$ using a combination
of {\em Planck}, SPT, and ACT anisotropy data, plus WMAP polarization~\cite{planck}.
More recently, BICEP2 CMB experiment has found an excess of $B$-mode power at degree angular scales,
indicating the presence of tensor modes with $r=0.20^{+0.07}_{-0.05}$ when assuming nearly scale-invariant tensor and scalar spectra~\cite{bicep2}. If this result is confirmed, it would give a very strong support to inflation model and open a new window
for probing the inflationary dynamics, though conflicting with the {\em Planck} low $r$ limit.

There have been many ideas trying to alleviate the tension on the high tensor-to-scalar ratio of BICEP2.  Here we will restrict ourself to those involving inflationary dynamics~\cite{suppression}. The main idea is to accommodate relatively large tensor contribution in the CMB anisotropy by suppressing the large-scale scalar spectrum with a transient fast-roll phase or sound-speed variation of inflaton fluctuations in a slow-roll inflation, or by anti-correlating the tensor and scalar modes for power cancellation. Recently, the authors in Ref.~\cite{mata,wu1} considered the effect on density perturbation due to a quantum environment that interacts with inflaton. It was shown that the quantum environment constitutes a colored noise that induces inflaton fluctuations, resulting in a suppression of the scalar spectrum at large-scales. This suppressed scalar power spectrum was then used to explain the observed low quadrupole in the CMB anisotropy. On the other hand, tensor modes generated during inflation are also affected by the quantum environment but only through gravitational interaction, so the effect is suppressed by the Planck mass and the tensor modes remain nearly intact. In light of this, we will show that by the same token an interacting inflaton could alleviate the tension between {\em Planck} and BICEP2 data.

\section{Colored noise}

Motivated by various theoretical reasons or cosmological observations, there has been a lot of interest in inflation models with an interacting inflaton such as reheating, preheating, or trapping phenomenon, in which the inflaton is coupled to scalar, fermion, or vector fields. Here we will not restrict ourselves to a specific inflation model, but rather assuming a successful inflation potential that satisfies the standard slow-roll condition. Furthermore, we introduce an interaction between the inflaton and scalar fields as a simple working example to show the occurrence of a colored noise due to the interaction during inflation and to study its effect on inflaton fluctuations. There are many studies on the effects of particle production during inflation, though in different contexts, deriving similar inflaton perturbation equations that exhibit both noise and dissipation effects~\cite{wu1,wu2,interactionpaper}.

The colored noise stems from the quantum interaction of the inflaton, $\phi$, and
other fields such as a scalar, $\sigma$, while the inflaton is rolling down the potential $V(\phi)$.
The Lagrangian of this kind of model usually takes the following form~(see, for examples, Refs.~\cite{wu1,wu2}),
\begin{eqnarray}
\mathcal{L}&=&\frac{1}{2}\partial^{\mu}\phi\,\partial_{\mu}\phi-V(\phi)+\mathcal{L}_\sigma\nonumber\\
  \mathcal{L}_\sigma  &=& \sum_i{1\over2}\left[ \partial^{\mu}\sigma_i\,\partial_{\mu}\sigma_i
             -m_{\sigma_i}^{2}\sigma_i^{2}-g_i^{2}(\phi-\bar{\phi}_i)^{2}\sigma_i^{2}\right],
\end{eqnarray}
where $g_i$ is a coupling constant and $\bar{\phi}_i$ is a constant field value. For a single $\sigma$ field and $\bar{\phi}=0$, it reduces to the simplest case in which a massive scalar couples to the inflaton~\cite{wu1}. In the case with many copies of $\sigma_i$ fields and $m_{\sigma_i}=0$, when $\phi$ rolls down to a trapping point $\bar{\phi}_i$, $\sigma_i$ particles become instantaneously massless and are copiously produced, and then backreact on the motion of the inflaton~\cite{wu2}. In either case, it has been shown that particle number density fluctuations (or the noise term) in the $\sigma_i$ particle production would induce a blue power spectrum of the inflaton fluctuations. Below we will give a brief review of this phenomenum. For the purpose of showing the generation of the blue power spectrum, we consider the simplest interaction in a small-field inflation model that has been investigated in Ref.~\cite{wu1}:
\begin{equation}
\mathcal{L}_\sigma= {1\over2}\partial^{\mu}\sigma\,\partial_{\mu}\sigma
             -{1\over2}m_{\sigma}^{2}\sigma^{2}-{1\over2}g^{2}\phi^{2}\sigma^{2}.
\end{equation}

We can approximate the background metric to be de~Sitter during inflation which is given by
\begin{equation}
ds^{2}=a^2 (\eta)(d\eta^{2}-d{\bf x}^2),
\end{equation}
where $a=-1/(H\eta)$ with the Hubble parameter $H$ and inflation begins at $a(\eta_i)=1$.
Then, using the influence functional method that integrates out the interaction term and the field $\sigma$,
and introduces an auxiliary field $\xi$, the effective action becomes~\cite{influence}
\begin{eqnarray}
S_{\rm eff} [\phi,\phi_{\Delta}, \xi ] && = \int d^4 x \, a^2 (\eta)
\, \phi_{\Delta} (x) \left\{ -\ddot{\phi}(x) - 2aH\dot{\phi}(x)
\right. \nonumber\\ & & \left. +\nabla^{2}\phi(x)  - a^{2}\left[
V'(\phi)
+g^{2}\langle\sigma^{2}\rangle\phi(x) \right] \right. \nonumber \\
&&-\left.   g^4 a^2 (\eta) \phi (x) \int d^{4} x' \, a^4(\eta') \,
\theta(\eta-\eta') \right.\nonumber\\ & & \times \left. i G_- (x,x')
\phi^2 (x') + g^2 a^2 (\eta) \phi (x) \xi(x) \right\} ,
\end{eqnarray}
where $\phi_{\Delta}$ is a relative field variable, the dot and prime denote respectively differentiation with respect to $\eta$ and $\phi$, and the kernels $G_{\pm}$ can be obtained from the Green's function of
$\sigma$:
\begin{equation}
G_{\pm}(x,x')=\langle\sigma(x)\sigma(x')\rangle^2 \pm
                               \langle\sigma(x')\sigma(x)\rangle^2.
\label{grfct}
\end{equation}
The effects from the quantum field $\sigma$ on the inflaton are given by the
dissipation via the kernel $G_{-}$ as well as a stochastic force
induced by the multiplicative colored noise $\xi$ with
\begin{equation}
\langle\xi(x)\xi(x')\rangle=  G_{+}(x,x').
\label{noise}
\end{equation}
Next, we extremize the effective action $ \delta S_{\rm eff}/
\delta \phi_{\Delta}$ and obtain the semiclassical Langevin equation
for $\phi$:
\begin{eqnarray}
&& \ddot{\phi}+2aH\dot{\phi}-\nabla^{2}\phi+a^{2}\left[V'(\phi)
+g^{2}\langle\sigma^{2}\rangle\phi\right]\nonumber\\&&
-g^{4}a^{2}{\phi} \int d^{4} x'
 a^4(\eta')
\theta(\eta-\eta')\,i\, G_{-}(x,x')
{\phi}^{2}(x')\nonumber\\&&\,\,\,\,\,\,=g^2 a^2 \, \phi \, \xi + \xi_w\, ,
\label{lange}
\end{eqnarray}
where we have included the white noise $\xi_w$ in the free-field stochastic
inflation~\cite{star} with
\begin{equation}
\langle\xi_w (x)\xi_w (x')\rangle\propto \delta (x-x'),
\label{wnoise}
\end{equation}
which can be produced by integrating out the high-frequency modes of the inflaton (see, for example, Ref.~\cite{mata}).
The white noise reproduces the inflaton vacuum quantum fluctuations
$\langle\varphi_q^{2}\rangle$ with a scale-invariant power spectrum
given by $\Delta^q_k=H^2/(4\pi^2)$~\cite{hawking}.
Note that the dissipation in the equation is not
important at the beginning of inflation because it is a time
accumulated term. We may safely neglect it if we only consider the first few efolds of inflation.
Let us drop the dissipation for the moment and consider the colored noise only. Then, after decomposing $\phi$ into a mean field and a classical
perturbation: $\phi(\eta,\bf x)={\bar\phi}(\eta) +
{\varphi}(\eta,\bf x)$, we obtain the linearized Langevin equation,
\begin{equation}
\ddot{\varphi}+2aH\dot{\varphi}-\nabla^{2}{\varphi}+ a^{2}
m_{\varphi{\rm eff}}^2 {\varphi} = g^2 a^2 {\bar\phi}\,
\xi,\label{varphieq}
\end{equation}
where the effective mass is $m_{\varphi{\rm
eff}}^2=V''({\bar\phi})+g^{2}\langle\sigma^{2}\rangle$ and the time
evolution of $\bar\phi$ is governed by $V({\bar\phi})$. The equation
of motion for $\sigma$ from which we construct its Green's function
can be read off from its quadratic terms in the Lagrangian as
\begin{equation}
\ddot{\sigma}+2aH\dot{\sigma}-\nabla^{2}{\sigma}+ a^{2} m_{\sigma}^2
{\sigma} = 0.
\label{sigmaeq}
\end{equation}
Let us decompose
\begin{eqnarray}
Y(x)&=&\int\frac{d^3{\bf k}}{(2\pi)^{3\over 2}} Y_{\bf
k}(\eta)\,e^{i{\bf k}\cdot{\bf x}}, \quad
{\rm where}\;Y=\varphi,\xi, \nonumber \\
\sigma(x)&=&\int\frac{d^3{\bf k}}{(2\pi)^{3\over 2}} \left[b_{\bf
k}\sigma_k(\eta)\,e^{i{\bf k}\cdot{\bf x}} + {\rm h.c.}\right],
\end{eqnarray}
where $b_{\bf k}^\dagger$ and $b_{\bf k}$ are creation and
annihilation operators satisfying $[b_{\bf k},b_{{\bf
k}'}^{\dagger}]= \delta({\bf k}-{\bf k}')$.
Then, the solution to Eq.~(\ref{varphieq}) is obtained as
\begin{eqnarray}
\varphi_{\bf k} (\eta)& = & -ig^2\int_{\eta_i}^{\eta} d\eta'
a^4(\eta') {\bar\phi}(\eta') \xi_{\bf k}(\eta')\nonumber\\
& &\,\,\,\times\left[\varphi_k^{1}(\eta')\varphi_k^{2}(\eta)
                   - \varphi_k^{2}(\eta')\varphi_k^{1}(\eta)\right],
\label{varphisol}
\end{eqnarray}
where the homogeneous solutions $\varphi_k^{1,2}$ are given by
\begin{equation}
\varphi_k^{1,2}={1\over2a} (\pi|\eta|)^{1\over 2}
                        H_\nu^{(1),(2)}(k\eta).
\end{equation}
Here $H_\nu^{(1)}$ and $H_\nu^{(2)}$ are Hankel functions of the
first and second kinds respectively and $\nu^2=9/4-m_{\varphi{\rm
eff}}^2/H^2$. In addition, we have from Eq.~(\ref{sigmaeq}) that
\begin{equation}
\sigma_{ k} ({\eta})={1\over2a} (\pi|\eta|)^{1\over 2} \left[c_1
H_\mu^{(1)}(k\eta)+c_2 H_\mu^{(2)}(k\eta)\right],
\end{equation}
where the constants $c_1$ and $c_2$ are subject to the normalization
condition, $|c_2|^2 -|c_1|^2=1$, and $\mu^2=9/4-m_{\sigma}^2/H^2$.

\begin{figure}[htbp]
\centerline{\psfig{file=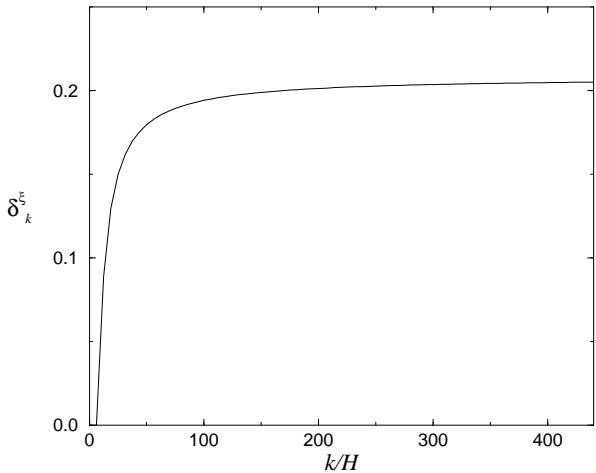, width=9cm}} \caption{Power
spectrum of the noise-driven inflaton fluctuations
$\delta^\xi_k\equiv 4\pi^2\Delta^\xi_k/g^4{\bar\phi}_0^2$, where
${\bar\phi}_0$ is the mean field near the beginning of inflation. The
starting point, $k/H=2\pi$, corresponds to the $k$-mode that leaves
the horizon at the start of inflation.} \label{ps}
\end{figure}

To maintain the slow-roll condition: $m_{\phi{\rm eff}}^2=m_{\varphi{\rm eff}}^2\ll H^2$ (i.e., $\nu=3/2$), requires
that $g^2<1$ and $m_{\sigma}^2>H^2$. The latter condition
limits the growth of $\langle\sigma^{2}\rangle$ during inflation to
be less than about $10^{-2}H^2$~\cite{ford,kkk}. Eq.~(\ref{sigmaeq}) does not include mass corrections to
$m_{\sigma}^2$ from the mean inflaton field, $g^{2}{\bar\phi}^2$,
and the mass renormalization due to quantum fluctuations of the inflaton,
$g^2\langle\varphi_q^{2}\rangle$.
Under the slow-roll condition, $\langle\varphi_q^{2}\rangle$ grows
linearly as $H^3t/4\pi^2$~\cite{ford,kkk} and thus
$\langle\varphi_q^{2}\rangle\simeq H^2$ after about $60$ efoldings
(i.e., $Ht\simeq 60$).  Therefore,
as long as $g^2{\bar\phi}^2 \le 2H^2$ for the period during which
those $k$ modes of cosmologically
relevant scales cross out the horizon, we can conveniently choose
$m_{\sigma}^2=2H^2$ (i.e., $\mu=1/2$) for which $\sigma$
takes a very simple form. After then, $g^2{\bar\phi}^2$ may grow to
a value much bigger than $H^2$ and thus the effective mass of $\sigma$
becomes much larger than $H^2$. If so, this large mass will suppress
the growth of $\langle\sigma^{2}\rangle$~\cite{kkk} and may diminish
the effect of the noise term. From now on, let us consider only the
relevant period with $g^2{\bar\phi}^2 \le 2H^2$.  It was shown that
when $\mu=1/2$ one can select the Bunch-Davies vacuum (i.e., $c_2=1$
and $c_1=0$)~\cite{kkk}. Hence, using Eqs.~(\ref{noise}) and
(\ref{varphisol}), we obtain~\cite{wu1}
\begin{equation}
\langle\varphi_{\bf k}(\eta)\varphi_{{\bf k}'}^*(\eta)\rangle
=\frac{2\pi^2}{k^3}\Delta^\xi_k(\eta) \delta({\bf k}-{\bf k}'),
\end{equation}
where the noise-driven power spectrum is given by
\begin{eqnarray}
\Delta^\xi_k(\eta)&=&\frac{g^4 z^2}{8\pi^4} \int_{z_i}^z dz_1
\int_{z_i}^z dz_2 \, {\bar\phi}(\eta_1) {\bar\phi}(\eta_2) \frac{\,
\sin z_{-}}{z_1 z_2 z_{-}}  \times \nonumber\\
&&\left[\sin(2\Lambda z_{-}/k)/z_{-}-1\right] F(z_1) F(z_2),
\label{pseq}
\end{eqnarray}
where $z_{-}=z_2-z_1$, $z=k\eta$, $z_i=k\eta_i=-k/H$, $\Lambda$
is the momentum cutoff introduced in the evaluation of the ultraviolet
divergent $k$-integration of $\sigma_k$ in the Green's
function~(\ref{grfct}), and
\begin{equation}
F(y)=\left(1+\frac{1}{yz}\right)\sin(y-z)+ \left({1\over
y}-{1\over z}\right)\cos(y-z).
\end{equation}
Note that the term $\sin(2\Lambda z_{-}/k)/z_{-} \simeq
\pi\delta(z_{-})$ when $\Lambda\gg k$, so $\Delta^\xi_k(\eta)$ is
insensitive to $\Lambda$. Both
${\bar\phi}(\eta_1)$ and ${\bar\phi}(\eta_2)$ in Eq.~(\ref{pseq})
can be approximated as a constant mean field ${\bar\phi}_0$.
$\Delta^\xi_k(\eta)$ at the horizon-crossing time given by $z=-2\pi$ was found in Ref.~\cite{wu1} and here we show the value versus $k/H$ in Fig.~\ref{ps}.
The figure shows that the noise-driven fluctuations
depend on the onset time of inflation and approach asymptotically to
a scale-invariant power spectrum $\Delta^\xi_k\simeq
0.2g^4{\bar\phi}_0^2/(4\pi^2)$ at large $k$. Here, the mean field value ${\bar\phi}_0$ should be naturally of order $H$ in small-field inflation.
In large-field inflation model discussed in Ref.~\cite{wu2}, the obtained noise-driven power spectrum is also blue and there ${\bar\phi}_0\sim H$ can be the spacing between the trapping points along the inflaton trajectory.

Let us go back to examine the dissipation term in the Langevin
equation~(\ref{lange}). As mentioned above, the dissipation is a time
accumulated effect. Near the beginning of inflation, by doing integration by parts of this term, it was found~\cite{wu1} that this
term only contributes a mass correction of about
$10^{-2}g^4{\bar\phi}_0^2$ to $m_{\varphi{\rm eff}}^2$, a small friction term of order $10^{-2}g^4{\bar\phi}_0^2a\dot{\phi}/H$
to Eq.~(\ref{lange}), and  a correction to the slope of the inflaton potential $V'(\phi)$ of order
$10^{-2}g^4{\bar\phi}_0^3$. All of these corrections can be neglected as long as $g^2{\bar\phi}_0^2\le 2H^2$.

One should emphasize that the white noise~(\ref{wnoise}) causes a delta-function response and thus produces a scale-invariant spectrum of inflaton quantum fluctuations, whereas the colored noise~(\ref{noise}) has a causal response due to an interaction between the inflaton and the quantum environment. The latter gives rise to a suppressed power spectrum at cosmologically relevant scales as long as inflation lasts for just about $60$ efoldings. Although the exact form of the colored noise varies with the type of interaction, the blue-tilted power spectrum is just a consequence of the causal response. This characteristic feature in the power spectrum should be quite general; it exists whenever there is interaction between the inflaton and any other quantum fields.

\section{Effects on large-scale CMB}

Here assuming a slow-roll inflation, the scalar power spectrum contains both inflaton quantum fluctutations and that driven by the colored noise, given by
\begin{equation}
P^S=P^{S}_{\Lambda CDM}\frac{1+\Delta^\xi_k/\Delta^q_k}{1+r'},
\end{equation}
where the noise contribution is quantified by a parameter
\begin{equation}
r'\equiv \lim_{k\to\infty}\Delta^\xi_k/\Delta^q_k\simeq 0.2g^4{\bar\phi}_0^2/H^2,
\end{equation}
the power spectrum of the inflaton quantum fluctutations is given by the scale-invariant $\Delta^q_k=H^2/(4\pi^2)$, and $P^{S}_{\Lambda CDM}$ is the scalar power spectrum in the {\em Planck} best-fit Lambda Cold Dark Model ($\Lambda$CDM)~\cite{planck}. In Fig.~\ref{ps}, we will need to specify the duration of inflation in order to determine the value of $k/H$ that corresponds to a given comoving scale. In the following, instead of fixing the duration of inflaton, we will treat the value of $k/H$ that corresponds to $0.05$Mpc$^{-1}$ as a free parameter denoted by $k_{0.05}$.

On the the hand, in order to fit BICEP2 data of the $BB$ power spectrum, we prepare a scale-invariant tensor power spectrum in the $\Lambda$CDM model, $P^T=P^{T}_{\Lambda CDM}$, such that the $r$ ratio, $P^T/P^{S}_{\Lambda CDM}=0.2$ at $k=0.002$Mpc$^{-1}$. Recently, {\em Planck} polarization data~\cite{planckdust} has implied that the BICEP2 B-mode signal may contain contributions from polarized dust emission, so the ratio $r$ may be reduced significantly. In light of this, we also study the case when $r=0.1$, in which we repeat every step as in the case with $r=0.2$ except using $r=0.1$ tensor power spectrum.

\begin{figure}[htbp]
\centerline{\psfig{file=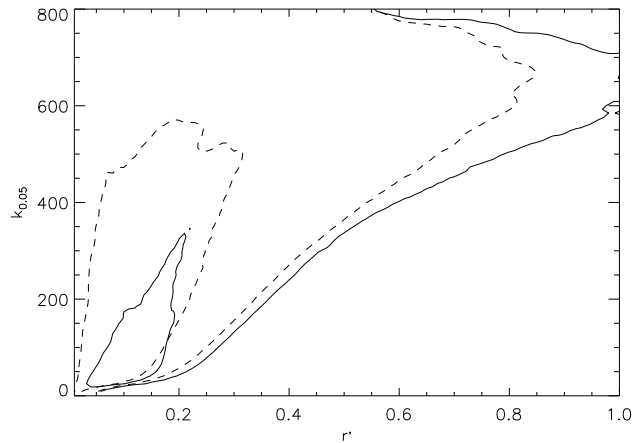, width=9cm}}
\caption{Solid lines are the likelihood plot of the parameters $r'$
and $k_{0.05}$ for the case with $r=0.2$, with $1$-sigma (the loop
in the left lower corner) and $2$-sigma (unbound) contours. The
maximum likelihood values are $r'=0.1$ and $k_{0.05}=33$. The dashed
lines are for the case with $r=0.1$, with the maximum likelihood
values given by $r'=0.05$ and $k_{0.05}=33$.} \label{contour}
\end{figure}

\begin{figure}[htbp]
\centerline{\psfig{file=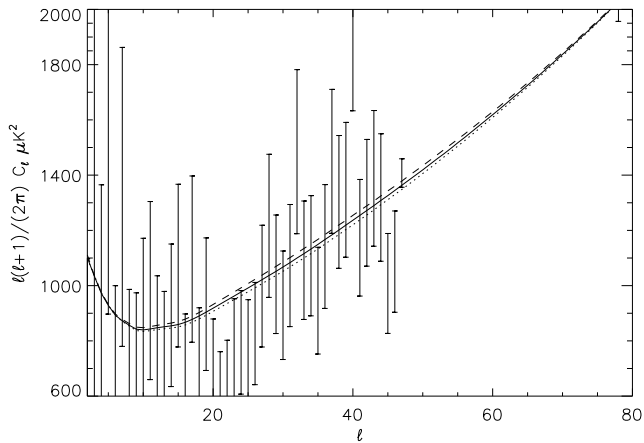,
width=9cm}} \caption{CMB temperature anisotropy power spectrum of
the colored noise model is denoted by the dashed (solid) line for
the case with $r=0.2$ ($r=0.1$). The dotted line is the power
spectrum predicted in the $\Lambda$CDM model. Overlaid are the {\em
Planck} measurements. For $l>70$, all three spectra overlap.}
\label{total}
\end{figure}

Then using $P^S$ and $P^T$ we compute CMB $TT$ and $BB$ power
spectra, based on the CMBFast code~\cite{SZ}. We tune the values of
$r'$ and $k_{0.05}$, by fixing the other cosmological parameters to
the best-fit values of the {\em Planck} $\Lambda$CDM
model~\cite{planck}, to best fit the {\em Planck} and BICEP2 data.
For the case with $r=0.2$, the likelihood plot in Fig.~\ref{contour} shows the maximum
likelihood values of $r'=0.1$ and $k_{0.05}=33$. Fig.~\ref{total}
shows that the $TT$ power spectrum with $r'=0.1$ and $k_{0.05}=33$,
which is induced by a combination of both $P^S$ and $P^T$, is indeed the same as
that induced only by $P^{S}_{\Lambda CDM}$ within measurement errors.
The large-scale $TT$ power suppression due to the colored noise can
be really made up by the tensor contribution. If ${\bar\phi}_0\sim H$ is assumed,
then $r'=0.1$ would imply that $g\sim 0.84$.
For the case with $r=0.1$, we find that
the results are very similar to those for $r=0.2$ except that the value of $r'$ is now
reduced to about $r'=0.05$. Interestingly, recent studies based on Bayesian statistics have indicated
that the {\em Planck} and BICEP2 joint likelihood analysis strongly favors a scalar power spectrum
like that one in Fig.~\ref{ps}~\cite{abazajian}.

\section{Conclusions}

Because inflaton has to be interacting with other fields so as to convert its potential energy into radiation to reheat the Universe, the interaction may induce a large-scale power suppression in the CMB temperature anisotropy power spectrum. We have shown that the presence of a significant amount of tensor modes as indicated in the BICEP2 measurement of the CMB $B$-mode polarization would lift up this temperature suppression. This implies that the measured temperature anisotropy power spectrum made by the {\em Planck} team can be a combination of scalar and $r=0.2$ tensor contributions. If BICEP2 results are confirmed to be genuine tensor $B$ modes, our work would indicate that we may have one-sigma detection of the interacting inflation model. On the other hand, if BICEP2 B-mode signal is mostly polarized dust contamination and $r$ is thus reduced significantly, this would put a tighter upper bound on the interaction strength. Lastly, we admit that the duration of inflation in the present work must be assumed to be about $60$ efoldings such that the power suppression takes place just at the large scales of the Universe.

This work was supported in part by the National Science Council,
Taiwan, ROC under the Grants No. NSC101-2112-M-001-010-MY3 (K.W.N.)
and NSC100-2112-M-032-001-MY3 (G.C.L. and I.C.W.).

\end{document}